\documentclass[aip,reprint,jcp,nolongbibliography]{revtex4-2}

\usepackage[]{placeins}

\usepackage{csquotes}
\usepackage{amsmath}
\usepackage{amssymb}
\usepackage{color}
\usepackage{bm}
\usepackage[normalem]{ulem}
\usepackage{hhline}
\usepackage{physics}
\usepackage{booktabs}
\usepackage{csquotes}
\usepackage{float}
\usepackage{xcolor,hyperref}
\hypersetup{
   colorlinks,
   linkcolor={blue!50!black},
   citecolor={blue!50!black},
   urlcolor={blue!80!black}
}
\usepackage{cleveref}

\usepackage{graphicx}

\DeclareMathAlphabet{\mathitbf}{OML}{cmm}{b}{it}

\newcommand{\dbar}{{\,\mathchar'26\mkern-12mu d}}

\newcommand{\Tp}{T_{\text{p}}}
\newcommand{\tfit}{\tilde{t}}
\newcommand{\lfit}{\tilde{\ell}}
\newcommand{\tbal}{t_{\text{ballistic}}}

\newcommand{\dyn}{\bm{\mathcal{M}}}
\newcommand{\Uihs}{U_{\text{IHS}}}

\begin{document}

\title{Does mesoscopic elasticity control viscous slowing down in glassforming liquids?}
\author{Geert Kapteijns}
\email{g.h.kapteijns@uva.nl}
\affiliation{Institute for Theoretical Physics, University of Amsterdam, Science Park 904, 1098 XH Amsterdam, The Netherlands}
\author{David Richard}
\affiliation{Institute for Theoretical Physics, University of Amsterdam, Science Park 904, 1098 XH Amsterdam, The Netherlands}
\author{Eran Bouchbinder}
\affiliation{Chemical and Biological Physics Department, Weizmann Institute of Science, Rehovot 7610001, Israel}
\author{Thomas B.~Schr{\o}der}
\affiliation{``Glass and Time", IMFUFA, Department of Science and Environment, Roskilde University, P.O.~Box 260, DK-4000 Roskilde, Denmark}
\author{Jeppe C. Dyre}
\affiliation{``Glass and Time", IMFUFA, Department of Science and Environment, Roskilde University, P.O.~Box 260, DK-4000 Roskilde, Denmark}
\author{Edan Lerner}
\affiliation{Institute for Theoretical Physics, University of Amsterdam, Science Park 904, 1098 XH Amsterdam, The Netherlands}

\begin{abstract}
\noindent The dramatic slowing down of relaxation dynamics of liquids approaching the glass transition remains a highly debated problem, where the crux of the puzzle resides in the elusive increase of the activation barrier $\Delta{E}(T)$ with decreasing temperature $T$. A class of theoretical frameworks --- known as \emph{elastic models} --- attribute this temperature dependence to the variations of the liquid's macroscopic elasticity, quantified by the high-frequency shear modulus $G_\infty(T)$. While elastic models find some support in a number of experimental studies, these models do not take into account the spatial structures, length scales, and heterogeneity associated with structural relaxation in supercooled liquids. Here, we propose and test the possibility that viscous slowing down is controlled by a \emph{mesoscopic} elastic stiffness $\kappa(T)$, defined as the characteristic stiffness of response fields to local dipole forces in the liquid's underlying inherent structures. First, we show that $\kappa(T)$ --- which is intimately related to the energy and length scales characterizing quasilocalized, nonphononic excitations in glasses --- increases more strongly with decreasing $T$ than the macroscopic inherent structure shear modulus $G(T)$ (the glass counterpart of liquids' $G_\infty(T)$) in several computer liquids. Second, we show that the simple relation $\Delta{E}(T)\!\propto\!\kappa(T)$ holds remarkably well for some computer liquids, suggesting a direct connection between the liquid's underlying mesoscopic elasticity and enthalpic energy barriers. On the other hand, we show that for other computer liquids, the above relation fails. Finally, we provide strong evidence that what distinguishes computer liquids in which the $\Delta{E}(T)\!\propto\!\kappa(T)$ relation holds, from those in which it does not, is that the latter feature highly fragmented/granular potential energy landscapes, where many sub-basins separated by low activation barriers exist. Under such conditions, it appears that the sub-basins do not properly represent the landscape properties relevant for structural relaxation. 
\end{abstract}

\maketitle

\section{Introduction}

Despite enormous research efforts over several decades, there is still no consensus regarding the mechanism that governs the stupendous increase in viscosity of supercooled liquids upon cooling toward their glass transition temperature $T_g$. \cite{shoving_2015,torchinsky_jcp_2009,MW_cates_length_discussion_prl_2017,Liesbeth2018,tarjus_can_explained_no_length_jcp_2019,turci_prx_2017,tanaka_order_parameter_nat_com_2019} That the relaxation time increases upon supercooling is not a mystery in itself; the challenge that this phenomenon presents is to explain why the liquid dynamics slows down so dramatically upon supercooling --- in most cases, the relaxation time or viscosity vary by 14 orders of magnitude over a reduction of less than 50\% in temperature.\cite{jeppe_review2006,Micoulaut_2016}

Solving the glass transition riddle amounts to understanding the temperature dependence of the activation energy $\Delta{E}(T)$ that enters a Arrhenius-like relation for the primary structural relaxation time $\tau_\alpha$ or the shear viscosity $\eta$ of a liquid, of the form
\begin{equation}\label{eq:kramers_relation}
\tau_\alpha(T) \sim \eta(T) \sim \exp\big(\Delta{E}(T)/k_{\mbox{\tiny B}}T\big)\,.
\end{equation}
The activation energy usually increases upon cooling, in a few cases stays nearly constant, and never decreases. One distinguishes between `fragile' and `strong' glassforming liquids by the \emph{relative} variation of the activation energy $\Delta{E}(T)$ throughout the relevant temperature range (approximately $[T_g,2T_g]$), with fragile (strong) systems featuring a larger (smaller) relative variation. The analysis of experimental and simulational data shows that in the most fragile systems the activation energy varies by nearly a factor of four across the relevant temperature range \cite{tarjus_2004,hecksher2008little}. Theoretical approaches that offer an explanation for the huge slowing-down of structural relaxation dynamics of supercooled liquids put forward throughout the years include Mode Coupling Theory,\cite{gotze2008complex,nandi2020microscopic} thermodynamic approaches,\cite{Kauzmann_1948,adam_gibbs} Random First Order Theory (RFOT),\cite{Wolynes_1989} Frustration-Limited Domains,\cite{Tarjus2005} locally-favored structures,\cite{Coslovich_2007_lfs,Hallett_2020} kinetically constrained frameworks,\cite{fredrickson1984kinetic, ritort2003glassy,Chandler_prx,elmatad2010finite} and many others described in several excellent reviews.\cite{heuer_review, Cavagna200951, paddy_huge_review_2015, charbonneau_review_2017, smarajit_review, Berthier_Biroli_RevModPhys_2011,speck2019dynamic,chandler2010dynamics}

Another set of approaches to the glass transition problem --- coined `elastic models'  --- attribute the slowing-down of supercooled liquids' dynamics to changes in their elastic properties that occur upon supercooling. \cite{elastic_model1996,elastic_model2004,elastic_model2006,jeppe_review2006,Larini2008,wyart_brito_2007,wyart_brito_2009,krausser2015interatomic,hasyim2021theory} A subset of elastic models link the activation energy to \emph{atomistic} elastic observables such as high frequency vibrational mechanics \cite{Larini2008} or the stability of soft modes pertaining to metastable states of the free energy.\cite{wyart_brito_2007,wyart_brito_2009}

Of relevance to our discussion here is a class of elastic models \cite{elastic_model1996,elastic_model2004,elastic_model2006,jeppe_review2006,shoving_2015} that adopts a continuum-elastic viewpoint. The so-called \emph{shoving model} proposes that relaxation events in supercooled liquids require an isotropic local volume expansion, regarding for simplicity the surrounding liquid as a homogeneous medium described by standard continuum elasticity theory. Since the associated strain field outside a local isotropic expansion is exclusively volume-conserving, shear-like in nature, \cite{landau_lifshitz_elasticity} and occurs over a short, elastic time scale, the shoving-model barrier toward structural relaxation is governed by the macroscopic high-frequency shear modulus $G_\infty$ (see the related discussion in Refs.~\citenum{flynn1968atomic, inst_shear_modulus_jeppe}), \emph{i.e.},
\begin{equation}\label{shoving_model}
\Delta{E}(T) \propto G_\infty(T)V_c\,.
\end{equation}
Here $V_c$ is a characteristic volume, which for simplicity is assumed to be temperature-independent. While some experiments support \autoref{shoving_model}, \cite{torchinsky_jcp_2009,shoving_2015} the shoving model lacks a concrete connection with the geometry, atomistic nature and mesoscopic spatial extent of relaxation events. In particular, the model does not take into account the spatially varying elastic properties of any supercooled liquid.

Which micromechanical observables do bear structural and geometrical similarities with structural relaxation events? Previous simulations \cite{Schober_correlate_modes_dynamics,widmer2008irreversible,harrowell_2009,schoenholz2014understanding} demonstrated that low-frequency, quasilocalized vibrational modes strongly correlate with the spatial organization and loci of relaxational flow events in supercooled liquids. Those observations suggest that the elastic stiffness relevant for structural relaxation events in supercooled liquids is that associated with soft, quasilocalized modes (QLMs) rather than macroscopic elastic moduli. Recently, there has been considerable progress in understanding the statistical, structural and energetic properties of QLMs in structural glasses. \cite{modes_prl_2016,ikeda_pnas,modes_prl_2018,LB_modes_2019,cge_paper,pinching_pnas,gonzalezlopez2020universality} It is now well-established that QLMs generically populate the low-frequency vibrational spectra of structural glasses. These modes' frequencies $\omega$ follow a universal distribution that grows from zero as $\omega^4$, independently of microscopic details,\cite{modes_prl_2016,gonzalezlopez2020universality} spatial dimension,\cite{modes_prl_2018} or glass preparation history, \cite{cge_paper,pinching_pnas,LB_modes_2019} in agreement with some previous theoretical predictions,\cite{soft_potential_model_1991,Gurevich2003} but at odds with others.\cite{Schirmacher_prl_2007,eric_boson_peak_emt,silvio,Ikeda_scattering_2018}

Since the energy distribution of soft QLMs is scale-free,\cite{modes_prl_2016,modes_prl_2018,gonzalezlopez2020universality} it is not possible to sharply determine their characteristic energy from the vibrational spectrum alone, as discussed at length in Refs.~\citenum{cge_paper,pinching_pnas}. While anharmonic frameworks that embed definitions of QLMs have been put forward,\cite{SciPost2016,kapteijns2020nonlinear,richard2021simple} it is currently impossible to detect the full population of soft excitations in a glass. Instead, in Refs.~\citenum{cge_paper,pinching_pnas} it was proposed that a good proxy for QLMs' characteristic energy can be obtained by considering the energy associated with a glass' typical response to local force dipoles, referred to in what follows as the \emph{dipole stiffness} $\kappa$. As we shall show below, $\kappa$ is highly sensitive to the formation history of the glass, and increases significantly more with annealing at low temperatures compared to the athermal shear modulus $G$.
\begin{figure}
    \centering
    \includegraphics[width=\columnwidth]{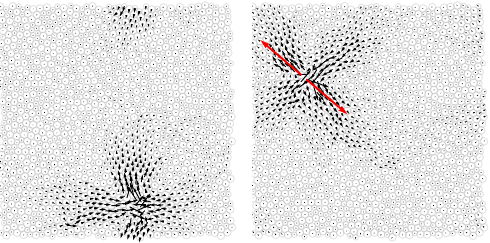}
    \caption{Structural similarity between a soft quasilocalized excitation (left) and a normalized response to a local force dipole (right) in a two-dimensional computer glass.\cite{cge_paper} The applied force dipole is shown in red.}
    \label{fig:mode_dipole_comparison}
\end{figure}
\begin{figure}
    \centering
    \includegraphics[width=\columnwidth]{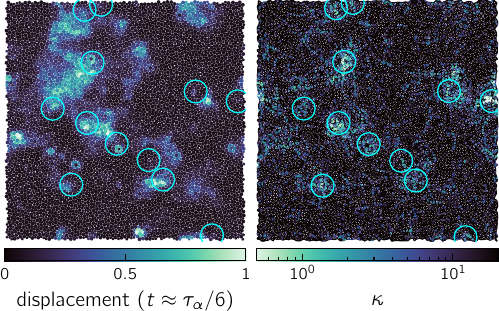}
    \caption{Propensity for motion \cite{harrowell_isoconfiguration} in the viscous liquid is highly correlated with the local dipole stiffness in a two-dimensional computer liquid.\cite{cge_paper} 
    (left) Isoconfigurational average of the neighbor-relative displacement \cite{mazoyer2009dynamics} after $t\!\approx\! \tau_{\alpha}/6$. (right) Local $\kappa$-field. The color of each bond represents the stiffness of the response to the force dipole between the two particles that constitute the bond, and bonds between all interacting pairs are shown (see \Cref{appendix:elastic_observables}). Circles denote the loci of soft QLMs, demonstrating that soft local dipole stiffnesses are correlated with soft QLMs (measured as described in \Cref{appendix:mode_extraction}).}
    \label{fig:microscopic_comparison}
\end{figure}
\begin{figure}
    \centering
    \includegraphics[width=\columnwidth]{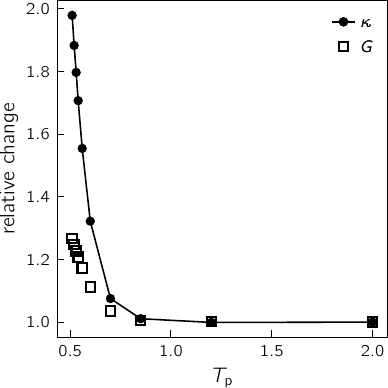}
    \caption{Relative change of the mean dipole stiffness $\kappa$ and athermal shear modulus $G$ in the 3DIPL model,\cite{cge_paper} as a function of the parent temperature $\Tp$ from which the glass samples were instantaneously quenched.}
    \label{fig:kappa_and_G}
\end{figure}

Let us briefly review and illustrate the key ideas of Refs.~\citenum{cge_paper,pinching_pnas} that are relevant to our discussion; \autoref{fig:mode_dipole_comparison} demonstrates the similarity between a soft QLM --- a \emph{rare} mechanical fluctuation --- and a \emph{typical} response of the same glass to a local force dipole. \autoref{fig:microscopic_comparison} shows that the spatial map of motion-propensity \cite{harrowell_isoconfiguration} in a viscous computer liquid is highly correlated with the spatial map of $\kappa$ measured in the inherent structure (IHS) underlying the initial liquid configuration. Importantly, \autoref{fig:microscopic_comparison} further shows (see circles and figure caption) that soft dipole stiffnesses are correlated with the loci of soft QLMs.\cite{cge_statistics_jcp_2020} Finally and of particular relevance here, \autoref{fig:kappa_and_G} demonstrates the sensitivity of the \emph{mean} dipole stiffness $\kappa(\Tp)$ to thermal annealing, quantified by the parent temperature $\Tp$ from which glasses were instantaneously quenched: the variation of $\kappa(\Tp)$ is almost 100\% over the range accessible with molecular dynamics simulations, compared to a variation of only 26\% of $G$.

The above line of reasoning --- \emph{i.e.}, that the characteristic stiffness of QLMs may control the structural relaxation rate, and that this highly annealing-dependent stiffness is faithfully captured by the characteristic (mean) dipole stiffness $\kappa(\Tp)$ --- naturally leads to the hypothesis that the energy scale $\Delta E(T)$ of \autoref{eq:kramers_relation} should be proportional to $\kappa(T)\!\equiv\!\kappa(\Tp = T)$. We emphasize that this hypothesis rests crucially on the simplifying assumption that upon lowering the temperature, the relevant energy barriers for relaxation in the viscous liquid are homogeneously rescaled by the single curvature $\kappa$, representing a characteristic energy scale of the glass.

In this work, we find that the relation $\Delta E(T)\!\propto\!\kappa(T)$ holds remarkably well in two canonical glassforming computer liquids (see details below), from the mildly supercooled liquid regime to the most deeply supercooled regime currently accessible with Graphical Processing Unit (GPU) simulations. These observations suggest that for some liquids there exists a direct connection between the liquid's underlying mesoscopic elasticity and energy barriers, and that the latter are enthalpic in nature. However, we also identify and study two other glassforming computer liquids for which $\Delta E(T)$ appears to grow faster than $\kappa(T)$ upon supercooling. We provide strong evidence that what distinguishes the two pairs of liquids is that the latter two feature a much more fragmented/granular potential energy landscape, characterized by many sub-basins separated by minute energy barriers. These are crossed at timescales much smaller than the typical \emph{vibrational} timescale of the supercooled liquid. Under such physical conditions, one does not expect $\kappa(T)$ --- which probes the mesoscopic elasticity of sub-basins --- to properly capture the landscape properties relevant for crossing the barrier $\Delta E(T)$ associated with slow structural relaxation.   

Finally, we perform a microscopic analysis of inherent structure mesoelastic properties (`mesoelastic' hereafter refers to mesoscopic elasticity), and show that, in models that follow $\Delta E(T)\!\propto\!\kappa(T)$, the \emph{local} $\kappa$ field correlates well with \emph{local} particle mobility, consistent with previous work.\cite{Schober_correlate_modes_dynamics,widmer2008irreversible,harrowell_2009,schoenholz2014understanding} These local, spatial correlations put the relation $\Delta E(T)\!\propto\!\kappa(T)$ on firmer, causal grounds. In contrast, we find that in models that do not obey $\Delta E(T)\!\propto\!\kappa(T)$, local correlations between the $\kappa$-field and local particle mobility deteriorate and disappear. 

The presentation of our results is organized as follows. In \autoref{sect:models_and_methods} we describe the computer glassforming models and physical observables studied in this work. In \autoref{sect:results} we present our results for all models, starting with a test of the hypothesis $\Delta{E}\!\propto\!\kappa(T)$, followed by an analysis of the complexity of the potential energy landscapes of the different models that reveals the origin of the observed differences between the models, and finally presenting a microscopic spatial-correlation analysis between mobility and the local stiffness. \autoref{sect:discussion} summarizes and discusses future research directions.

\section{Models and methods}
\label{sect:models_and_methods}

\subsection{Models}
We employ four different glassforming models in three dimensions. We describe each model briefly, and refer the reader elsewhere for detailed descriptions. 

\noindent\emph{Inverse-power-law (IPL)} --- The IPL model is a 50:50 binary mixture of particles interacting with a pairwise potential $U(r)\!\sim\! r^{-10}$.\cite{cge_paper} We also employ the two-dimensional version of this model for illustrative purposes in \autoref{fig:mode_dipole_comparison} and \autoref{fig:microscopic_comparison}. 

\noindent\emph{Modified binary Lennard-Jones (mBLJ)} --- This model is a modification of the Kob-Andersen 80:20 binary mixture \cite{kablj} that is more resistant to crystallization due to the decreased strength of the AA and BB interactions. \cite{solid_msd_thomas_jeppe_jcp_2020}

\noindent\emph{Sticky Spheres (SS)} --- The SS model is a 50:50 binary mixture with a pairwise potential that is piecewise defined; the repulsive part is identical to that of the Lennard-Jones potential, whereas the attractive part is much stronger and is cut off at a shorter range.\cite{zylberg_sticky_spheres_pre_2011} We use the model with a dimensionless cutoff $x_c\!=\! 2^{1/6}\!\times\!1.2$ referred to as the CSS model in Ref.~\citenum{karina_sticky1}.

\noindent\emph{Stillinger-Weber (SW)} --- The SW model \cite{Stillinger_Weber} is a network glassformer with a three-body term in the potential energy that favors local tetrahedral geometry. We chose all parameters equal to the original silicon model, except for the strength of the three-body term, which we put to $\lambda = 18.75$. This choice was found to optimize glassforming ability.\cite{russo2018glass} A detailed description, including the expression for the Hessian matrix (defined in~\Cref{appendix:elastic_observables}), may be found in Ref.~\citenum{gonzalezlopez2020universality}.

\subsection{Simulation details}

Simulations were performed\footnote{We employed the molecular dynamics software packages RUMD \cite{RUMD} for the mBLJ model, HOOMD-blue \cite{HOOMD} for the IPL and SS models, and LAMMPS \cite{LAMMPS} for the SW model.} in the $NVT$-ensemble using the Nosé–Hoover thermostat. \cite{nose_hoover_jcp_1984} The density $\rho$, system size $N$, integration time step $\delta t$, and Nosé-Hoover coupling time $\tau_{\text{NH}}$ for each model are given in \autoref{tab:rho}. At each temperature, we thermalized the liquid until the mean squared displacement (MSD) and the self-part of the intermediate scattering function  \cite{binder2011glassy} $F_{\text{s}}(t; q_{\max})$ became time-translationally invariant.\footnote{All correlators were calculated with the Atooms-Postprocessing package \cite{atooms_pp} and the RUMD analysis tools.\cite{RUMD}} The correlators $F_{\text{s}}$ and MSD are always understood to pertain to the largest particle species, and $q_{\max}$ corresponds to the position of the first peak of the static structure factor of the largest particle species, unless noted otherwise. The value of $q_{\text{max}}$ for each model is given in \autoref{tab:rho}.

After thermalization, we performed equilibrium measurements and regularly performed instantaneous quenches using either the FIRE algorithm \cite{fire} or conjugate-gradient minimization to create ensembles of inherent structures, which are labeled by the parent temperature $T_p$ from which they were quenched. We define the structural relaxation time $\tau_{\alpha}$ according to $F_{\text{s}}(\tau_{\alpha}; q_{\text{max}}) = 0.2$. The elastic quantities $G$ and $\kappa$ are calculated in the ensembles of inherent structures; this procedure, including the statistics needed to obtain ensemble averages, is described in \Cref{appendix:elastic_observables}.

\begin{table}[h]
\begin{tabular}{@{}lllll@{}}
\toprule
       & IPL  & mBLJ & SS  & SW\footnote{For a conversion to SI units based on the original parametrization of crystalline silicon, see Ref.~\citenum{Stillinger_Weber}.}  \\ \midrule
$\rho$ & 0.82 & 1.2  & 0.6 & 0.52 \\
$N$  & 2000 & 8000 & 3000 & 8000   \\
$\delta t$ & 0.005 & 0.005 & 0.005 & 0.065 \\
$\tau_{\text{NH}}$ & 0.2 & 0.2 & 0.2 & 6.5 \\
$q_{\text{max}}$  & 6.35 & 7.25 & 5.47 & 6.28 \\ \bottomrule
\end{tabular}
\caption{Density $\rho$, system size $N$, integration time step $\delta t$, coupling time of the Nosé-Hoover thermostat $\tau_{\text{NH}}$, and the position of the first peak of the static structure factor of the largest particle species $q_{\max}$ in microscopic units for all models studied.}
\label{tab:rho}
\end{table}

\subsection{Units}
\label{subsec:units}

All particles in all models have mass $m$, and we work in units in which $m\!=\!k_{\text{B}}\!=\!1$. 
To facilitate comparison between models, all quantities are reported in units of length $a_0\!\equiv\!(V/N)^{1/3}$ and time $t_0\!\equiv\! a_0/\sqrt{T}$ (except temperature and density, which are unity in these units, and are reported in microscopic units instead).
This choice of units guarantees a collapse of the MSD in the regime of ballistic motion, which ends at approximately $t/t_0\!\approx\! 0.1$ for all models, see \Cref{appendix:short_time_msd}.

\section{Results}\label{sect:results}

\subsection{The relation between activation energy and the characteristic dipole stiffness in supercooled liquids}\label{subsect:kappa_hypothesis}

\begin{figure*}
    \centering
    \includegraphics[width=\textwidth]{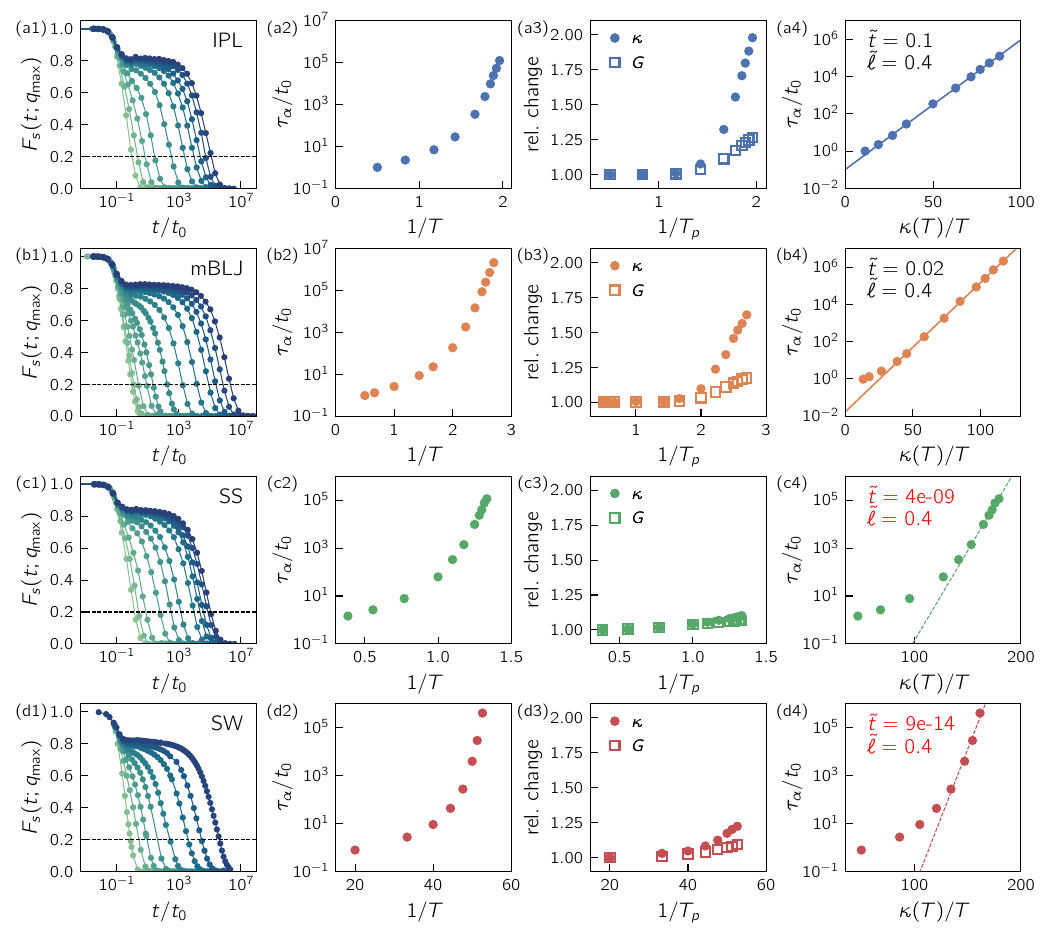}
    \caption{Structural relaxation in the supercooled liquid is controlled by $\kappa(T)$ --- the mean stiffness of the underlying glass' response to local force dipoles --- in the IPL and mBLJ models, but \emph{not} in the SS and SW models. (a1-d1) Self-part of the intermediate scattering function $F_{\text{s}}(t; q_{\text{max}})$ \emph{vs.} $t/t_0$. The structural relaxation time $\tau_{\alpha}$ is defined as $F(\tau_{\alpha}; q_{\max}) = 0.2$; the threshold 0.2 is shown with a dashed line. (a2-d2) $\tau_{\alpha}/t_0$ \emph{vs.}~$1/T$. (a3-d3) Relative change of $\kappa$ and $G$ as a function of $1/\Tp$; for each model, the relative increase of $\kappa$ is greater than that of $G$.
    (a4-d4) Fit to the mesoelastic model \autoref{eq:arrhenius_fit} [where $\tfit$ and  $\lfit$ are defined according to $\log(\tau_{\alpha}/t_0)\!=\!(\lfit a_0)^2 \kappa(T)/T+\log\tfit$], demonstrating that the activation energy of the supercooled liquid is given by $\kappa$ calculated in the glass in the IPL and mBLJ models (row one and two), but not in the SS and SW models (row three and four).}
    \label{fig:tau_vs_kappa_over_T_all}
\end{figure*}

We start by testing the proposition that the activation energy of supercooled liquids is proportional to the characteristic glass dipole stiffness $\kappa(T)$. Specifically, we test a functional relation of the form
\begin{equation}
\label{eq:arrhenius_fit}
     \log\!\left(\frac{\tau_{\alpha}}{t_0}\right) \sim \frac{\kappa(T)}{T} \ ,
\end{equation}
which is hereafter referred to as the \emph{mesoelastic model}. We present results for all four glassforming models side-by-side in \autoref{fig:tau_vs_kappa_over_T_all}. Panels (a1-d1) show the self-part of the intermediate scattering function for all temperatures studied, from which the structural relaxation is extracted as $F_{\text{s}}(\tau_{\alpha}; q_{\text{max}}) = 0.2$ (see \autoref{sect:models_and_methods}). Panels (a2-d2) show the structural relaxation time $\tau_{\alpha}$ \emph{vs.}~$1/T$. Panels (a3-d3) show the relative change of $\kappa$ and $G$ as a function $1/\Tp$, demonstrating that $\kappa$ is more sensitive to annealing than $G$ for all models studied (see \autoref{fig:kappa_and_G}).

Finally, panels (a4-d4) of \autoref{fig:tau_vs_kappa_over_T_all}  show the fit to \autoref{eq:arrhenius_fit}, and constitute the main result of this paper. Remarkably, for the IPL and mBLJ models, the functional form is valid from the very start of the activated regime --- around $t/t_0\!\approx\!10$, which is around $100$ times the duration of the ballistic regime of the MSD (see \Cref{appendix:short_time_msd}) --- demonstrating that the activation energy in these supercooled liquids is determined by $\kappa(T)$. This observation suggests that the activation energy barrier $\Delta{E}(T)$ is enthalpic in nature, as no entropic considerations have been invoked. The fit to \autoref{eq:arrhenius_fit} involves the dimensionless time and length $(\tfit, \lfit)$, defined according to $\log(\tau_{\alpha}/t_0)\!=\!(\lfit a_0)^2 \kappa(T)/T+\log\tfit$. We find that $(\tfit, \lfit)$ are $(0.1, 0.4)$ for the IPL model and $(0.02, 0.4)$ for the mBLJ model, comparable in magnitude to the physical reduced scales of the duration of the ballistic regime and the particle diameter.

\begin{figure*}
    \centering
    \includegraphics[width=\textwidth]{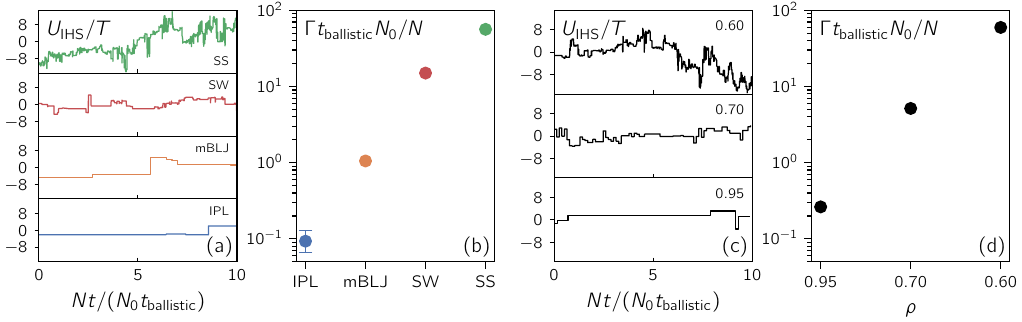}
    \caption{The potential energy landscape of the SS and SW models is more fragmented/granular than that of the mBLJ and IPL models. (a) Inherent structure potential energy $\Uihs$ (with respect to an arbitrary reference value) rescaled by the temperature $T$, \emph{vs.}~reduced time $Nt/(N_0 \tbal)$ (see discussion in text). (b) Reduced inherent structure transition rate. Panels (c) and (d) show the same analysis for the SS model at three different densities, demonstrating that the fragmentation/granularity of the potential energy landscape decreases substantially with increasing density in this computer liquid. Error bars represent the 90\% confidence interval obtained with the Bootstrap method \cite{bootstrap} (using the software package scikits-bootstrap \cite{bootstrap_package}). Except for the IPL model, error bars are smaller than the symbol size.}
    \label{fig:hopping} 
\end{figure*}

Intriguingly, in the cases of the SS and SW systems, the mesoelastic model of \autoref{eq:arrhenius_fit} breaks down --- the tentative range of validity is too small or nonexistent, and the tentative value of $\tfit$ is many orders of magnitude smaller than any reduced timescale intrinsic to the system. What is the origin of the breakdown of the mesoelastic model in some cases and not in others? In the next Subsection, we propose that dramatic changes in the properties of the potential energy landscape (PEL) between the different glassformers studied here are responsible for the observed breakdown of the mesoelastic model in the SS and SW systems.

\subsection{Fragmentation/Granularity of the potential energy landscape}
\label{subsect:granularity}

What distinguishes the respective PELs of the IPL and mBLJ models, whose relaxational dynamics are well-predicted by the mesoelastic model, from the SS and SW models, for which the mesoelastic model fails? Here we show that the potential energy landscape of the SS and SW models is much more \emph{fragmented/granular}, meaning that the configuration-space volume (basin) associated (\emph{e.g.}, by steepest descent dynamics~\cite{Stillinger1983}) with any single inherent structure (local minimum of the PEL) is much smaller. As a result, when considering the inherent structure trajectories that underlie a supercooled liquid dynamical trajectory, one observes very frequent transitions between nearby minima on the PEL, over timescales much smaller than any of the system's intrinsic timescales. We note that this phenomenon was already observed in Ref.~\citenum{giovambattista2002transitions} for the extended simple point charge model of water.\cite{berendsen1987missing} To perform a quantitative assessment of the PEL fragmentation/granularity of the glass forming models under consideration, we inspect short-time inherent structure trajectories, as done \emph{e.g.}~in Ref.~\citenum{landscape_dominated_jeppe_2000}. To meaningfully compare between different models, we used state points of comparably high viscosity, see further details in \Cref{appendix:similar_viscosity}. 

In \autoref{fig:hopping}(a), a representative trajectory of the inherent structure potential energy $\Uihs$ (relative to an arbitrary reference value) rescaled by the temperature $T$ is plotted \emph{vs.}~the reduced time $N t / (N_0 \tbal)$ for each model. We have made energy and time dimensionless to be able to assess all models on the same footing. The unit of time $\tbal N_0/N$ is explained as follows. First, $\tbal$ represents the duration of the regime of ballistic motion, which is approximately $0.1 t_0$ for all models (see \Cref{appendix:short_time_msd}). Second, since transitions on the potential energy landscape consist of \emph{local} molecular rearrangements, the number of inherent structure jumps per unit time is expected to scale with the system size $N$.\cite{heuer_review}
We set $N_0\!=\!1000$ arbitrarily, so that by counting the number of jumps in $\Uihs$ in this unit of time, we are effectively measuring the number of transitions a system of 1000 particles would make per ballistic timescale $\tbal$. Note that we measure the \emph{extensive} potential energy differences, since the potential energy difference between subsequent inherent structures is expected to be $N$-independent.\cite{heuer_review}

\begin{figure*}[]
    \centering
    \includegraphics[]{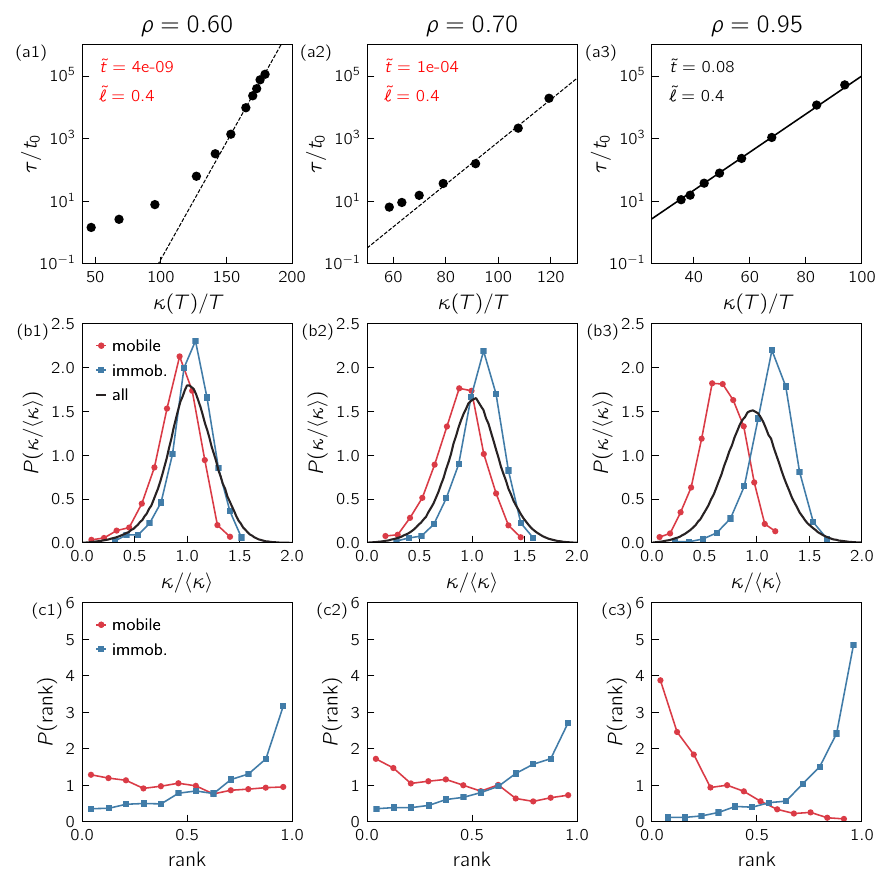}
    \caption{The local correlation between dipole stiffness $\kappa$ and mobility is high in the SS model at high density, but decreases significantly as the density is lowered. (a1-3) Fit to the mesoelastic model of \autoref{eq:arrhenius_fit} (cf.~\autoref{fig:tau_vs_kappa_over_T_all}). (b1-3) Distribution of local $\kappa$ rescaled by the bulk average $P(\kappa/\expval{\kappa})$ for the mobile, immobile and total population (see definitions in text). (c1-3) Rank distribution of local $\kappa$ for the mobile and immobile population (0 and 1 represent the lowest and highest value of the configuration).}
    \label{fig:microscopic} 
\end{figure*}

The contrast between different models is striking; in the IPL and mBLJ models, the system resides in a single inherent structure for a substantial period of time. In the SS and SW models however, the system makes transitions at a rate far higher than the inverse of any intrinsic microscopic timescale. To quantify this observation, we plot the reduced inherent structure transition rate $\Gamma \tbal N_0/N$ for each model in \autoref{fig:hopping}(b), which was averaged over ten trajectories with different initial equilibrium configurations.\footnote{In principle, back-and-forth inherent structure transitions can occur in arbitrary small time windows $\Delta t$, causing an undercount of the true number of transitions. Furthermore, we omit from our count transitions that occur between two states with identical $\Uihs$, since we measure transitions as jumps $\Delta \Uihs > \epsilon$ with $\epsilon = 10^{-5}$ (we verified that changing $\epsilon$ over several orders of magnitude does not change the results). We chose an integration time step equal to the molecular dynamics time step for the IPL and mBLJ models, and equal to \emph{one-tenth} the molecular dynamics time step for the SS and SW models, so that the undercounting effect of back-and-forth transitions is expected to be small. The undercounting effect of transitions with $\Delta \Uihs = 0$ is also expected to be small.} There is at least an order of magnitude difference between the transition rates of the two pairs of models. Remarkably, the IPL and mBLJ models are also separated by an order of magnitude difference, indicating that in the mBLJ model --- where a 1000-particle system at a viscous state point transitions between different basins of the PEL once every ballistic time --- the potential energy landscape is already appreciably more intricate than in the IPL model. We include movies of representative inherent structure trajectories for all models in the Supplementary Material.\textbf{[Ref.~to be inserted by editor]}

Can the fragmentation/granularity of a glassformer's PEL be altered by varying a \emph{single} external parameter? To address this question, we consider the SS model, and study its dynamics and landscape granularity under variations of the density $\rho$. We select the three densities 0.6, 0.7 and 0.95 (see \Cref{appendix:SS_details} for simulation details), and focus our analysis on state points of similarly high viscosity (see \Cref{appendix:similar_viscosity}). We note that across all three densities, we observe no difference in the degree of dynamical heterogeneity, see \autoref{fig:msd_collapse_different_densities}(c) in \Cref{appendix:similar_viscosity}. In \autoref{fig:hopping}(c) we present characteristic inherent structure potential energy trajectories for supercooled dynamics of the SS model at the aforementioned densities. As with the four models discussed above, we find that the inherent structure trajectory becomes increasingly less erratic as the density is increased, signaling an associated reduction of the degree of PEL granularity. In \autoref{fig:hopping}(d) we report the PEL interbasin transition rate; we find that increasing $\rho$ in the SS model results in a decrease of this rate by over two decades, similar to the variation observed between the four glassformers discussed above.

The dramatic decrease upon compression of the PEL interbasin transition rate in the SS model suggests that mesoelastic properties of the glasses underlying the denser states should be predictive of those states' supercooled relaxation dynamics. This suggestion is corroborated in \autoref{fig:microscopic}(a1-3), in which we show that the mesoelastic model is predictive of the dense [$\rho\!=\!0.95$, panel (a3)] SS liquid dynamics. However, as expected, it fails as the density is decreased [see panels (a1) and (a2)], as the liquids' PEL becomes more granular.

\subsection{Local correlations between dipole stiffness and mobility}

Until this point, we have presented and discussed evidence for a relation --- or lack thereof --- between the \emph{mean} mesoscopic dipole stiffness $\kappa$ and the \emph{mean} primary structural relaxation time $\tau_\alpha$. In order to establish that this observed relation between mesoscopic elasticity and dynamics is \emph{causal}, in this final section we investigate the correlation between \emph{local} dipole stiffness fields and \emph{local} fields of particle mobility. We focus on the SS model under variations of the density $\rho$, as explained above. In \autoref{subsect:granularity} we pointed out that, at $\rho\!=\!0.6$ (`low density'), the mesoelastic model of \autoref{eq:arrhenius_fit} is not predictive, and hence we expect there to be weak or no correlation between local dipole stiffness fields and local particle mobility; at sufficiently high densities, we have seen that the dynamics \emph{are} predicted by the mesoelastic model, and therefore expect to find local correlations between the $\kappa$-field and particle mobility.

To test these expectations, we define the particle mobility, or propensity of motion, as the particle's average displacement in the isoconfigurational ensemble \cite{harrowell_isoconfiguration} after a time $\tau_{\alpha}/10$. For each density, we collected 240 independent inherent structures from which we started 20 independent simulations with velocities drawn from the Maxwell-Boltzmann distribution. For each initial configuration, we then determined the most mobile and least mobile particles of the large species\footnote{All results in this section hold also for the small particle species. Correlations involving mobility have to be calculated on a per-species basis however, because the expected mobility of small particles is higher than that of large particles.} in each spherical region of radius $r_{\text{cut}}^{BB}$ (the cutoff radius of the large-large interaction\footnote{We confirmed that the results are not dependend upon reasonable changes in this cutoff length.}), and selected from this subset the 5 most mobile and 5 least mobile particles. With this procedure, we thus obtain a population of $240\!\times\!5\!=\!1200$ large particles at the mobile and immobile extremes. We define the local $\kappa_i$ for particle $i$ as the arithmetic mean of $\kappa_{ij}$ (see \autoref{eq:kappa_ij}) over the set of interacting pairs $ij$. In \autoref{fig:microscopic}(b1-3), we report the distribution of local $\kappa$ (rescaled by the bulk average) $P(\kappa/\!\expval{\kappa})$ for the mobile, immobile and total population. As expected, for the highest density [see \autoref{fig:microscopic}(b3)] the mobile and immobile distributions are well separated, with mobile and immobile particles featuring significantly below- and above-average $\kappa$, respectively. Upon lowering the density [see \autoref{fig:microscopic}(b1-2)], the separation decreases and finally almost vanishes.

To quantify the degree of correlation, for each initial inherent structure we rank the local $\kappa$-field on a per-species basis, with 0 and 1 representing the lowest and highest values. \autoref{fig:microscopic}(c1-3) shows the distribution of ranks of the mobile and immobile population for each density. For the highest density, the most mobile and immobile particles are overwhelmingly also the particles with respectively the lowest and highest $\kappa$. In accordance with the data in panels (c1-3), at intermediate and low density this correlation becomes significantly smaller. It is noteworthy however, that even for the lowest density the \emph{immobile} particles have a relatively high probability to have a very large $\kappa$, whereas the correlation between high mobility and small $\kappa$ breaks down almost completely. We further comment on this observation in \autoref{sect:discussion}.

\begin{figure}
    \centering
    \includegraphics[width=\columnwidth]{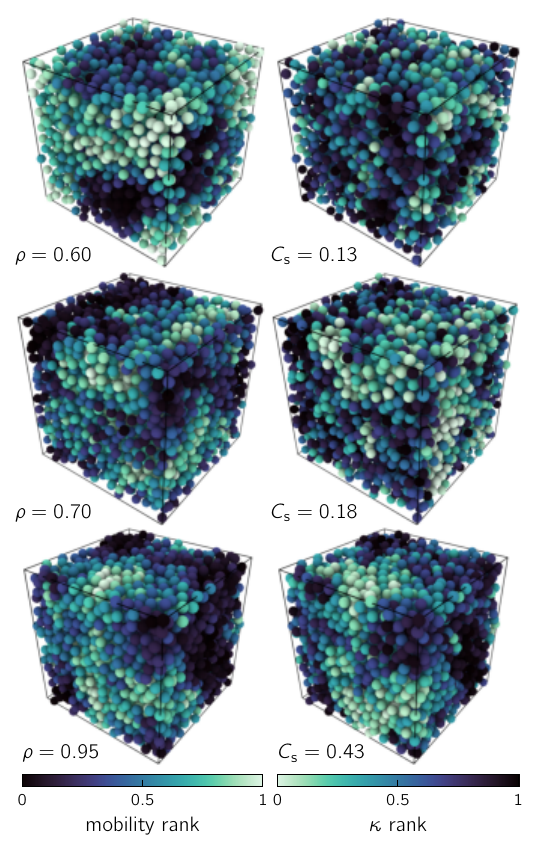}
    \caption{Local $\kappa$ and mobility rankings for a representative configuration at each density $\rho$, demonstrating that that the correlation $C_{\text{s}}$ decreases significantly as the glass's PEL granularity is increased by lowering the density. See text for definitions and additional discussion.}
    \label{fig:3d_snapshots} 
\end{figure}

To visualize these correlations, in \autoref{fig:3d_snapshots} we show side-by-side particle maps of the $\kappa$ and mobility rankings for a representative configuration at each density. Finally, we calculate Spearman's rank correlation coefficient $C_{\text{s}}$ --- the correlation coefficient between the rankings of the inverse of $\kappa$ and the mobility --- of the total particle population on a per-species basis. For each density, we report the ensemble average of the species-averaged $C_{\text{s}}$ in \autoref{fig:3d_snapshots}, once again confirming that the degree of correlation decreases significantly as the density is lowered.\footnote{We note that the reported values for $C_{\text{s}}$ increase if the local $\kappa$ and mobility fields are spatially coarse-grained, as was done in Ref.~\citenum{tanaka_order_parameter_nat_com_2019}, but that the decreasing trend with decreasing density remains the same.}

\section{Discussion}
\label{sect:discussion}
\begin{figure}
    \centering
    \includegraphics[width=\columnwidth]{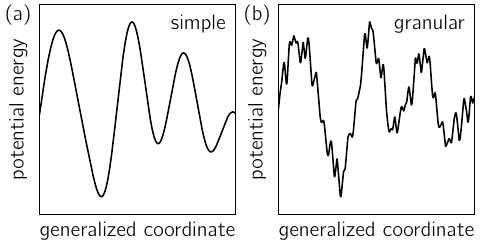}
    \caption{The cartoons illustrate the difference between the PELs of `simple' glass forming models [panel (a), schematically representing \emph{e.g.}~the IPL and mBLJ models], whose supercooled dynamics appear to be controlled by mesoelastic properties, and highly granular PELs [panel (b), schematically representing \emph{e.g.}~the SW and SS models], associated with models in which dynamics appear to be indifferent to underlying inherent structure elastic properties. Similar illustrations can be found in previous literature.\cite{Samwer_2014,MW_cates_length_discussion_prl_2017,karina_sticky2,heuer_review}}
    \label{fig:landscape_cartoon} 
\end{figure}

In this work, we put forward and tested the hypothesis that the activation energy $\Delta E$ that controls structural relaxation in supercooled liquids is proportional to the characteristic stiffness scale $\kappa$ of the underlying glass' response to local force dipoles. We find that the proposed relation $\Delta E\! \propto\!\kappa$ holds remarkably well in two canonical glassforming models (IPL and mBLJ), which also suggests that activation barriers against structural relaxation in these systems are predominantly enthalpic in nature.\cite{jeppe_review2006,MW_cates_length_discussion_prl_2017} This suggestion stands at odds with recent claims of Ref.~\citenum{baityjesi2021revisiting}, in which it is argued that the activation barriers in a similar glassformer are predominantly entropic, at least in the moderately supercooled regime. Further research is needed to reconcile the results of this work with ours.

We further highlighted the existence of correlations between local $\kappa$-fields in inherent structures that underlie equilibrium states and particle mobility in the ancestral supercooled liquid. These correlations reinforce the causality of the observed relation between bulk mesoscopic elasticity and supercooled liquid dynamics. We also show that $\Delta E\! \propto\!\kappa$ breaks down for two other well studied glassforming models, namely the SS and SW models. We have shown that this breakdown correlates with a dramatic increase in the degree of \emph{fragmentation/granularity} of these respective models' PELs, as illustrated by the cartoon in \autoref{fig:landscape_cartoon}. Finally, we established that on the local (particle) level, the spatial correlation between low-$\kappa$ regions --- indicative of the loci of soft quasilocalized excitations --- and high-mobility regions gradually deteriorates as the degree of PEL fragmentation/granularity is increased. 

We stress that we tested a hypothesis that may be deemed as excessively simple; inspired by the shoving model,\cite{jeppe_review2006} we posited that upon lowering the temperature, the relevant energy barriers for relaxation in the viscous liquid are proportional to the characteristic curvature $\kappa$ of QLMs. We agree with Goldstein's assertion in his seminal paper on the potential energy barrier picture of viscous liquids \cite{Goldstein1969} that \enquote{[t]here is every reason to think that the viscosity of a liquid is a property sensitive to fine details of the laws of force between the individual molecules of the liquid, as well as to the presence or absence of internal molecular degrees of freedom. No theory \textelp{} with a major stress laid on one or another thermodynamic property, itself a gross average over all the complexities of molecular interaction and shape, is going to be anything better than semiquantitative in character}. He added, however, that there exists \enquote{an extremely useful role for crude theories of viscosity: those that are found to be successful may suggest directions for more theoretically oriented and rigorous approaches to follow.} Judged by this standard, we have exceeded expectations by identifying a bulk-average quantity that \emph{quantitatively} predicts the structural relaxation over the entire (simulationally accessible) viscous range in two commonly studied glassformers.

Our results for models that obey $\Delta E \propto \kappa$ reinforce the hypothesis --- established by the early findings of Refs.~\citenum{Schober_correlate_modes_dynamics,widmer2008irreversible,harrowell_2009,schoenholz2014understanding} --- that mesoscopic elasticity and soft quasilocalized excitations should play an important role in future theories of viscosity, dynamical heterogeneity and facilitation. This idea is strengthened by the work of Ref.~\citenum{landes2021elastoplasticity}, which demonstrates that dynamical facilitation is elastically mediated. Combined, these results suggest QLMs' mutual interaction may be the underlying mechanism of dynamical facilitation, in analogy to how QLMs' collective dynamics determine plastic flow under shear deformation.\cite{lemaitre2004, manning2011, plastic_modes_prerc, nicolas2018deformation}
These important questions underline the necessity of reliably extracting the full field of soft QLMs (a study that is currently underway). 
Furthermore, it would be interesting to correlate local particle mobility with micromechanical properties of soft QLMs such as their stiffness and anharmonicity of their associated local energies,\cite{kapteijns2020nonlinear} especially under shear. Finally, the connection between mesoscopic elasticity and structural relaxation should be verified out of equilibrium (below $T_g$).

Our results for the models that do \emph{not} obey $\Delta E \propto \kappa$ make clear that the connection between soft excitations in the inherent structure and structural relaxation in the supercooled liquid is not universal. In fact, these models' extremely high transition rate between inherent structures raises doubts about the assumption that a single inherent structure is representative of the glass phase. Future research should aim to understand the nature of these rapid transitions, and determine if a distinction can be made between transitions that contribute to relaxation and those that do not.

An important future research direction is the determination of the causal factors affecting the degree of granularity of glassformers' PELs, and a more robust theoretical connection between PEL granularity and liquid dynamics.
As a first step, we demonstrate a clear correlation between high PEL fragmentation/granularity and dominance of attractive forces, as made evident by overlapping our models' pair correlation functions with their pairwise potentials, see \autoref{fig:gr} of \Cref{appendix:correlation_pel_attraction}.
This correlation raises the question whether in the models where attractive forces dominate, there exists a separate bond-breaking energy scale that may cause the correlation between mesoscopic elasticity and viscous dynamics to break down. We note, however, that the introduction of a bond-breaking energy scale has been shown to lead to strong (\emph{i.e.}~Arrhenius) dynamic behavior --- in contrast to the models where $\Delta E \sim \kappa$ fails, which remain highly fragile.
A further question is whether it is possible to relate a liquid's PEL granularity to mechanical inannealability\cite{karina_sticky2} --- the strong suppression of low-temperature thermal annealing-induced stiffening of $G$ and $\kappa$ [as demonstrated in \autoref{fig:tau_vs_kappa_over_T_all}(c3, d3)] --- since both phenomena seem to emerge in systems featuring strong attractive interactions. 
Such interactions have also been shown to induce the stiffening and depletion of QLMs, accompanied by changes in QLMs' destabilization mechanisms and in their spatial structure and geometry.\cite{karina_sticky1} Future research will have to determine why the connection between soft QLMs and structural relaxation can break down under the influence of attractive interactions, and what role, if any, soft QLMs play in structural relaxation in this class of systems.

Another important topic for future study is the relation between soft QLMs, the stiffness $\kappa$ and structural relaxation near the unjamming transition, \cite{ohern2003,liu_review,van_hecke_review} which marks the point at which purely repulsive systems lose their rigidity as the degree of connectedness of the underlying network of strong interactions is reduced below a critical value. Near this transition, the potential energy landscape becomes hierarchical and complex, and inherent structure transitions can be both localized and delocalized, \cite{scalliet2019nature} suggesting that the simple picture $\Delta E \sim \kappa$ might not be valid in this case. Interestingly, both repulsive liquids near the unjamming transition and low-density liquids with strong attractive interactions feature a complex energy landscape; the differences and similarities between these two situations warrants further research. Furthermore, the limit of extreme stickiness is characterized by unusual, non-two-step relaxation in the viscous regime, \cite{fullerton2020glassy} which presents another striking similarity to the unjamming transition (we emphasize that the attraction-dominated models studied in this paper do not approach this extremely sticky limit, as they feature conventional two-step decay of correlation functions).

A further promising future research direction is the connection between the stiffness $\kappa$ and ``softness", a machine-learned quantity that represents the distance to the hyperplane that optimally separates mobile and immobile particles in a high-dimensional space of local-structure-descriptors.\cite{machine_learning_1, machine_learning_2, sharp2018machine} The probability for a particle to rearrange has been shown to be Arrhenius in the softness, and the structural relaxation time to be a function of the average softness of the system,\cite{schoenholz2017relationship} both in conventional models and in a network glassformer.\cite{cubuk2020unifying} It would be especially interesting to use the softness to assess the predictive power of inherent structure configurations in systems with a high inter-inherent-structure transition rate. 

We have shown that even in the models with highly fragmented PELs, immobile particles in supercooled states correlate well with regions of large local $\kappa$-fields, even though mobile particles do \emph{not} correlate with low-$\kappa$ regions. It would therefore be natural to search for correlations between high-$\kappa$ regions in the underlying glasses and locally favored structures (LFSs). Finally, in Ref.~\citenum{glen_prl_2014} the authors argued that the correlation between LFSs with mobility is strongly system-dependent and typically very poor for constituents interacting with softer repulsion potentials (\emph{e.g.}, harmonic spheres). Thus, determining how the $\kappa$-field correlates with mobility in this class of glassformers is a logical next step.


\section*{Supplementary Material}

The supplementary videos show a representative inherent structure trajectory for each of the four main models studied in the paper.

\acknowledgements

Discussions with Ulf Pedersen, Corrado Rainone and Massimo Pica Ciamarra are warmly acknowledged. D.~R.~acknowledges support of the Simons Foundation for the ``Cracking the Glass Problem Collaboration" Award No.~348126. E.~L.~acknowledges support from the NWO (Vidi grant no.~680-47-554/3259). E.~B.~acknowledges support from the Minerva Foundation with funding from the Federal German Ministry for Education and Research, the Ben May Center for Chemical Theory and Computation, and the Harold Perlman Family. This work was supported in part by the VILLUM Foundation's \textit{Matter} grant (No. 16515).

\section*{Data availability}
The data that support the findings of this study are available from the corresponding author upon reasonable request.

\section*{Appendices}
\appendix

\section{Elastic observables}
\label{appendix:elastic_observables}

\subsection{Definitions}
We measure elastic properties in ensembles of inherent structures that underlie equilibrium liquid configurations, and generally depend on the parent (equilibrium) temperature $T_{\text{p}}$ from which they were quenched. 
We focus on two observables: the macroscopic athermal shear modulus $G$, defined as \cite{lutsko,athermal_elasticity}
\begin{equation}\label{athermal_G_definition}
\begin{split}
G &\equiv \frac{1}{V}\eval{\pdv[2]{U}{\gamma}}_{\vb*{f}=0,\gamma=0} \\
&=  \frac{1}{V} \eval{\left( \pdv[2]{U}{\gamma} - \pdv{U}{\gamma}{\vb*{x}} \cdot \dyn^{-1} \cdot \pdv{U}{\gamma}{\vb*{x}}\right)}_{\gamma=0} \\
&= \frac{1}{V} \eval{ \left( \pdv{U}{\vb*{\epsilon}_{yy}} + \pdv[2]{U}{\vb*{\epsilon}_{xy}} \right)}_{\vb*{f}=0,\vb*{\epsilon}=0},
\end{split}
\end{equation}
where $\vb*{x}$ are the particles' coordinates, $\dyn \equiv \eval{\pdv{U}{\vb*{x}}{\vb*{x}}}_{\vb*{x}=\vb*{x}_0}$ is the Hessian matrix of the potential energy in the inherent structure $\vb*{x}_0$, $\gamma$ is a strain parameter that enters the imposed affine deformation $\vb*{x} \to \vb*{H}(\gamma)\vb*{x}$ with $\vb*{H}(\gamma) = \vb*{I} + \gamma\, \vu{i} \otimes \vu{j}$, and $\vb*{\epsilon} \equiv \frac{1}{2} (\vb*{H}^T\vb*{H} - \vb*{I})$ is the strain tensor. We emphasize that in the second and fourth part of the equation the derivatives are taken under the constraint that the net forces remain zero, and $\eval{\left( \dots \right)}_{\gamma=0}$ means that the expression is evaluated at $\gamma=0$ (and similarly for $\eval{\left( \dots \right)}_{\vb*{\epsilon}=0}$). Here and in what follows, $\dyn^{-1}$ denotes the inverse of the Hessian matrix in the $(N-1)\dbar$-dimensional positive-definite subspace that excludes the $\dbar$ translational zero modes. This notation is justified because $\dyn^{-1}$ is always applied to a translation-free vector. Equations of the form $\vb*{x} = \dyn^{-1}\vb*{y}$ are solved with the Conjugate Gradient Method.\cite{barrett1994templates}

The second elastic observable we focus on is the typical stiffness $\kappa$ associated with the mesoscopic response of a glass to a local force dipole (see \autoref{fig:mode_dipole_comparison}). The normalized dipolar force $\vu*{d}^{ij}$ acting on the $i,j$ pair of interacting particles is 
\begin{equation}
\vu*{d}^{ij} \equiv \frac{1}{\sqrt{2}}\pdv{r_{ij}}{\vb*{x}}.
\end{equation}
The linear response of a glass to this force is
\begin{equation}\label{eq:z_ij}
\vb*{z}^{ij} = \dyn^{-1} \cdot \vu*{d}^{ij}.
\end{equation}
The stiffness associated with the response $\vb*{z}^{ij}$ is then
\begin{equation}\label{eq:kappa_ij}
\kappa_{ij} = \frac{ \vb*{z}^{ij} \cdot \dyn \cdot \vb*{z}^{ij}}{\vb*{z}^{ij}\cdot\vb*{z}^{ij}} = \frac{\vu*{d}^{ij} \cdot \dyn^{-1} \cdot \vu*{d}^{ij}}{\vu*{d}^{ij} \cdot \dyn^{-2} \cdot \vu*{d}^{ij}}.
\end{equation}
We note that the form of \autoref{eq:kappa_ij} implies that $\kappa_{ij}$ is sensitive to nearby soft QLMs, as described in Ref.~\citenum{cge_statistics_jcp_2020}.

\subsection{Ensembles of inherent structures}
Throughout the main text, we denote by $\kappa(T_{\text{p}})$ the average of stiffnesses $\kappa_{ij}$ (\autoref{eq:kappa_ij}) over randomly selected interactions $i,j$ in each glass of the ensemble with parent temperature $T_{\text{p}}$. In \autoref{tab:ensemble_size}, we list for each model the minimum ensemble size and number of dipole stiffnesses $\kappa_{ij}$ calculated per temperature. In some models we used significantly larger ensembles --- and therefore calculated significantly more dipole stiffnesses --- for the higher temperatures, than listed in \autoref{tab:ensemble_size}. This was required because the distribution $P(\kappa_{ij})$ features a power-law tail for small $\kappa_{ij}$, related to the universal distribution of soft quasilocalized excitations $D(\omega) \sim \omega^4$, \cite{cge_statistics_jcp_2020} the prefactor of which can be significantly higher for poorly annealed ensembles, and varies from model to model.\cite{karina_sticky1, karina_sticky2} In each case, we validated that the 90\% confidence interval calculated with the Bootstrap method \cite{bootstrap} (using the software package scikits-bootstrap \cite{bootstrap_package}) was smaller than the symbol size used in \autoref{fig:tau_vs_kappa_over_T_all}.   

We denote by $G(T_{\text{p}})$ the ensemble average of the athermal shear modulus as defined in \autoref{athermal_G_definition}. 

\begin{table}[]
\begin{tabular}{@{}lllllll@{}}
\toprule
Model  & IPL   & mBLJ  & SW    &   & SS &   \\ \cmidrule{5-7}
Density &    &  &  & 0.6  & 0.7 & 0.95 \\ \midrule
Ensemble size      & 10000 & 128   & 200    & 238 & 49 & 93 \\
\# dipole stiffnesses        & 2M    & 164K    & 80K  & 476K  & 98K & 186K \\ \bottomrule
\end{tabular}
\caption{Ensemble size and number of dipole stiffnesses $\kappa_{ij}$ (\autoref{eq:kappa_ij}) calculated per parent temperature, for each model.}
\label{tab:ensemble_size}
\end{table}

\section{Extraction of soft quasilocalized excitations}\label{appendix:mode_extraction}

In \autoref{fig:microscopic_comparison} we show the loci of soft QLMs in the 2DIPL model \cite{cge_paper} at a viscous state point ($T = 0.48$). It is currently not possible to extract the full population of soft QLMs, but we are able to find a subset with the following method: for each interacting pair of particles $ij$ we calculate the dipole response $\vu*{z}^{ij}$ (\autoref{eq:z_ij}), which we use as an initial condition to minimize the cost function described in Ref.~\citenum{richard2021simple} to obtain a \emph{pseudoharmonic mode} $\vu*{\pi}$. A pseudoharmonic mode accurately represents quasilocalized excitations, since it does not hybridize with phononic modes.\cite{kapteijns2020nonlinear} We show the twelve excitations with the smallest stiffness $\lambda(\vu*{\pi}) = \vu*{\pi} \cdot \dyn \cdot \vu*{\pi}$.

\section{Simulation details of the SS model at different densities}\label{appendix:SS_details}

In \autoref{tab:SS_simulation_details} we report the simulation details for the NVT simulations of the SS model at the three different densities $0.6$, $0.7$ and $0.95$.

\begin{table}
\begin{tabular}{llll} 
\toprule
$\rho$            & 0.6 & 0.7 & 0.95  \\\midrule
$N$               & 3000 & 4000  & 4000 \\
$\delta t$               & 0.005  & 0.003  & 0.002 \\
$\tau_{\text{NH}}$          & 0.2 & 0.2 & 0.2 \\
$q_{\text{max}}$          & 5.47 & 5.87 & 6.40 \\
\bottomrule
\end{tabular}
\caption{System size $N$, integration time step $\delta t$, coupling time of the Nosé-Hoover thermostat $\tau_{\text{NH}}$, and the position of the first peak of the static structure factor of the largest particle species $q_{\max}$ in microscopic units for the SS model at various densities $\rho$.}\label{tab:SS_simulation_details}
\end{table}

\begin{figure*}
    \centering
    \includegraphics[width=\textwidth]{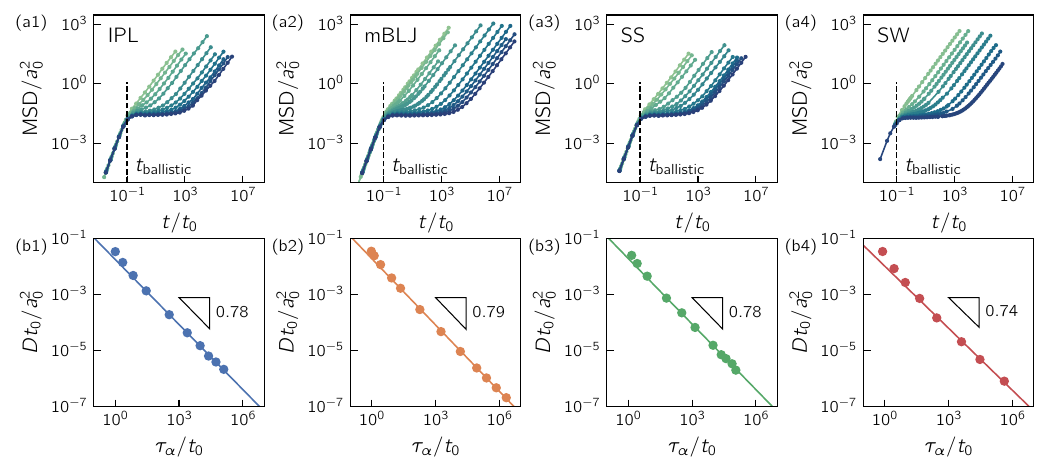}
    \caption{(a1-a4) Mean squared displacement (MSD) \emph{vs.}~time in reduced units $a_0$ and $t_0$ (see text for definitions). In all models, the ballistic regime ends approximately at $0.1t_0$. (b1-b4) Diffusion coefficient $D$ \emph{vs.}~relaxation time $\tau_{\alpha}$ in reduced units, demonstrating Stokes-Einstein breakdown with an exponent in the range 0.74-0.79 for all models.}
    \label{fig:msd_all}
\end{figure*}

\begin{figure}
    \centering
    \includegraphics[width=\columnwidth]{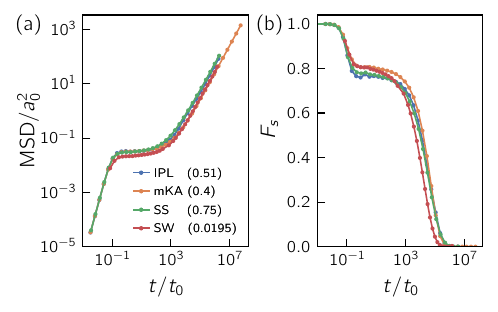}
    \caption{State points of similar viscosity across models, as determined by a collapse in the species-averaged MSD [panel (a)] and $F_s$ [panel (b)] in reduced units. 
    The deeply supercooled temperature we selected for each model is given in the legend of panel (a).}
    \label{fig:msd_collapse_different_models}
\end{figure}

\begin{figure*}
    \centering
    \includegraphics{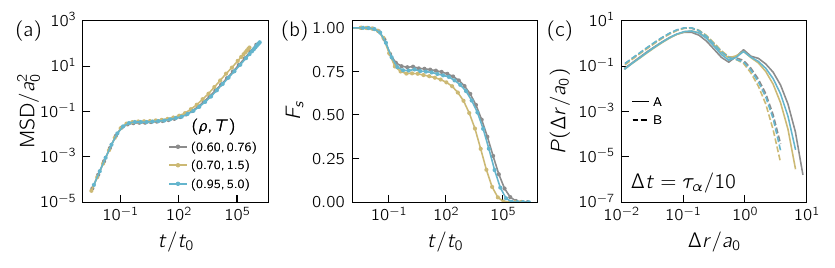}
    \caption{State points of similar viscosity in the SS model at different densities, as determined by a collapse in the species-averaged MSD [panel (a)] and $F_s$ [panel (b)] in reduced units. 
    The deeply supercooled state points we selected are given in the legend of panel (a). Panel (c) shows that the distribution of particle displacements after a time interval $\Delta t = \tau_{\alpha}/10$ is almost identical across densities.}
    \label{fig:msd_collapse_different_densities}
\end{figure*}

\section{Additional data}\label{appendix:additional_data}

\begin{figure*}
    \centering
    \includegraphics[width=\textwidth]{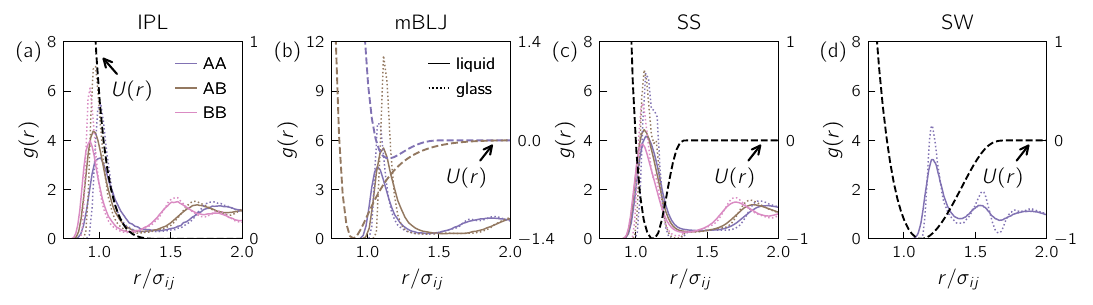}
    \caption{Interactions are dominated by hard-core repulsion in the IPL and mBLJ models, whereas in the SS and SW models, attractive interactions dominate. Solid lines denote the radial distribution function $g(r)$ in the liquid, dotted lines in the glass (left y-axes). The dashed lines denote the pairwise potential (right y-axes).
    In the mBLJ model, the potential is not a single function of $r/\sigma_{ij}$ for all pairs of species; therefore we only show the AA- and AB-distributions (accounting for 64\% and 32\% of the total radial distibution function, respectively). The SW model has only a single particle species.}
    \label{fig:gr}
\end{figure*}

\subsection{Short-time collapse of the MSD}\label{appendix:short_time_msd}
Rescaling distance and time by $a_0$ and $t_0$ trivially guarantees that in equilibrium, $\text{MSD} / a_0^2 = 3 (t / t_0)^2$ in the regime of ballistic motion (in units where $m = k_{\text{B}} = 1$). In \autoref{fig:msd_all}(a1-4) we show that for all models, the ballistic regime ends around $0.1 t_0$, which we refer to as $\tbal$. In \autoref{fig:msd_all}(b1-4) we show the breakdown of the Stokes-Einstein relation \cite{harris2009fractional, hiss1973diffusion} by plotting the rescaled diffusion coefficient $D t_0 / a_0^2$ \emph{vs.} $\tau_{\alpha} / t_0$. The breakdown exponent is in the range 0.74-0.79 for all models.

\subsection{Selection of state points of similar viscosity across models}\label{appendix:similar_viscosity}

In order to compare high-viscosity state points across models, we selected for each model a deeply supercooled temperature so that the species-averaged mean squared displacement and self-intermediate scattering function in reduced units approximately collapse. We report this collapse in \autoref{fig:msd_collapse_different_models}(a, b).

We follow the same procedure in selecting state points of similar viscosity in the SS model at different densities, the result of which is shown in \autoref{fig:msd_collapse_different_densities}(a, b). Additionally, \autoref{fig:msd_collapse_different_densities}(c) shows the distribution of particle displacements (in reduced units) after a time interval $\Delta t = \tau_{\alpha}/10$, demonstrating that the dynamical heterogeneity is almost completely insensitive to density. 

\subsection{Correlation between PEL fragmentation/granularity and presence of attractive interactions}
\label{appendix:correlation_pel_attraction}


In this Appendix, we show that in the models that obey the relation $\Delta E \propto \kappa$ and feature a generic/nongranular PEL (IPL and mBLJ), hard-core repulsive interactions are dominant; in contrast, in the models that do \emph{not} obey the relation $\Delta E \propto \kappa$ and feature a highly granular PEL (SS and SW), attractive interactions are dominant.

To demonstrate this, in \autoref{fig:gr} we plot the radial distribution function $g(r)$, and overlay the pairwise potential $U(r)$ for all models. For each model, we selected a state point of comparably high viscosity to perform the $g(r)$ measurement, see \Cref{appendix:similar_viscosity}.
In the IPL model, shown in panel (a), all interactions are by definition repulsive.
The mBLJ model, shown in panel (b), \emph{does} feature an attractive part. At the simulated density, the most common AA interaction (accounting for 64\% of the weight of the total radial distribution function) is dominantly repulsive, whereas the less common AB interaction (accounting for 32\% of the total) is dominantly attractive. For the SS and SW models shown in panels (b) and (c), the situation is very different. In the SS model, a significant portion of interactions utilize the very strong attractive part of the potential. In the SW model, this is the case for virtually \emph{all} interactions.

\FloatBarrier
\bibliography{references_jeppe_project}

\end{document}